# MAGNETIC PROXIMITY EFFECTS IN V/Fe SUPERCONDUCTOR/FERROMAGNET SINGLE BILAYER REVEALED BY WAVEGUIDE-ENHANCED POLARIZED NEUTRON REFLECTOMETRY

Yu.N. Khaydukov[1], V.L. Aksenov[1], Yu.V. Nikitenko[1], K.N. Zhernenkov[1,2], B. Nagy[3], A. Teichert[4], R. Steitz[4], L. Bottyán[3]

[1] Joint Institute for Nuclear Research, Dubna, Russia

[2] Ruhr-Universität Bochum, Bochum, Germany

[3] KFKI Research Institute for Particle and Nuclear Physics, Budapest, Hungary

[4] Helmholtz-Zentrum für Materialien und Energie, Berlin, Germany

ABSTRACT

Polarized neutron reflectometry is used to study the magnetic proximity effect in a superconductor/ferromagnet (SC/FM) system of composition Cu(32nm)/V(40nm)/Fe(1nm)/MgO. In contrast to previous studies, here a single SC/FM bilayer, is studied and multilayer artefacts are excluded. The necessary signal enhancement is achieved by waveguide resonance, i.e. preparing the V(40nm)/Fe(1nm) SC/FM bilayer sandwiched by the highly reflective MgO substrate and Cu top layer, respectively . A new magnetic state of the system was observed at temperatures below 0.7 $T_C$. manifested in a systematic change in the height and width of the waveguide resonance peak. Upon increasing the temperature from 0.7 $T_C$ to $T_C$, a gradual decay of this state is observed, accompanied by a 5% growth of the diffuse scattering. According to theoretical studies, such behavior is the result of the magnetic proximity effect. Due to the presence of the thin FM layer the superconducting electrons are polarized and, as a result, near the SC/FM interface an additional magnetic layer appears in the SC with thickness comparable to $\xi$, the coherence length of the superconductor.

INTRODUCTION

The term "proximity effect" was introduced in condensed matter physics in the 1960s considering the contact of a superconductor and a normal metal and was meant for the appearance of the superconducting order parameter in the normal metal on a length scale of $\xi$, the coherence length of the superconductor. In the early 2000s several theoretical publications [1,2] dealt with another type of proximity effect in hybrid superconductor/ferromagnet (SC/FM) heterostructures, namely the appearance of the magnetic order parameter in the superconductor close to the SC/FM interface. The effect was named magnetic [1], or inverse [2] proximity effect. Penetration of the magnetic order also occurs on a length scale of $\xi$ which, for conventional SCs is typically 10-100 nm. Thus, these effects are long-range effects. The sign of the induced magnetization depends on the quality of the SC/FM interface. In [1] and [2] the transport through the SC/FM interface was considered in the diffusive limit, i.e for the case of highly disordered (rough, dirty, terrace-rich) interfaces. The induced magnetization in the diffusive limit points antiparallel to the magnetization of the FM layer [3]. On the contrary, for a smooth (clean, terrace-free, etc.) SC/FM interface the ballistic limit applies and a parallel orientation of the FM and the induced magnetization is expected [3]. Therefore the quality of the SC/FM interface plays a crucial role in the considered problem. The temperature dependence of the magnetic proximity effect is investigated in several papers [2,4,5]. In all cases the proximity is manifested with maximum effects at temperatures far below $T_C$, the superconducting transition temperature. Upon rising the temperature to $T_C$, the induced magnetization gradually disappears.

Experimental observation of the magnetic proximity effect requires methods which, on the one hand are of high sensitivity and, on the other hand, provide a measure of the distribution of the magnetic moment within the SC/FM heterostructure with a resolution much less than $\xi$. One such method, as we shall see in the next section, is polarized neutron reflectometry (PNR). The method has already been used to study the magnetic state of SC/FM systems. In [6] the V(36.5nm)/Fe$_{0.5}$V$_{0.5}$(4.7nm)/[V(4.7nm)/Fe(4.7nm)]$_{10}$/MgO multilayer system was investigated. In this complex structure, the thick (36.5nm) vanadium layer is the SC layer in contact with the alloyed Fe$_{0.5}$V$_{0.5}$ FM layer of 4.7 nm thickness. The periodic structure [V(4.7nm)/Fe(4.7nm)]$_{10}$ was used to generate neutron standing waves in order to increase the depth-selectivity of PNR [7]. The authors reported a rather complex magnetic profile in the entire system below $T_C$, i.e. the superconducting transition of the thick vanadium layer influenced the magnetic profile of the entire structure, not only of the neighboring FM layer. In [8] and [9] a [La$_{2/3}$Ca$_{1/3}$MnO$_3$(10nm)/YBa$_2$Cu$_3$O$_7$(10nm)]$_{6-7}$ multilayer was investigated in a wide temperature range at various pressures. The investigated samples consisted of 6 or 7 bilayers of LaCaMnO FM, and high-$T_C$ YBaCuO SC layers. The doubling of the magnetic period, compared to the structural one was found when the temperature dropped below $T_C$. A similar period doubling was observed in [10] for the V(39 nm)/Fe(3 nm)/[V(3nm)/Fe(3nm)]$_{20}$/MgO periodic structure. The magnetic period doubling can be attributed to an RKKY-type interaction of the neighboring FM layers through the SC spacer [11].

In the above the PNR signal was enhanced to a detectable level by using multilayer samples. However, the large number of SC/FM interfaces seems to have complicated the interpretation and even hindered the actual effect. First of all, superconductivity exists only in layers the thickness of which is of order of $\xi$ or above [12,13,14]. Preparation of periodic SC/FM structures with SC layers of such thickness may inevitably lead to an increased cumulative roughness of the SC/FM interfaces, consequently to a distribution of the transparency of the various interfaces. Another problem is the above-mentioned RKKY-like interaction of the neighboring interfaces. In summary, the [SC/FM]$_n$ multilayer exhibits a more complex behavior, than a plain set of $n$ independent SC/FM interfaces. We believe, therefore, that proximity effects can be more reliably studied on two- and three-layered SC/FM multilayers. For example a three-layer SC/FM/SC system was studied by PNR in [15]. An anomalously large decrease (around 90% from the initial value) of the FM magnetization was found below the superconducting temperature. Such a decrease can be equally due to the decrease of saturation magnetization and a change in the domain state. Distinction could have been made following a quantitative analysis of the diffuse scattering, which, in that study, was unfortunately of too low intensity to be conclusive.

In the present work we report on proximity studies of a single SC/FM bilayer with composition V(40nm)/Fe(1nm). In order to enhance the weak magnetic scattering we used the waveguide enhancement of the neutron standing waves [16]. Using this regime, at temperatures $T < 0.7T_C$, we were able to explore a new magnetic state of the bilayer, the appearance of which we attribute to the magnetic proximity effect.

POLARIZED NEUTRON REFLECTOMETRY

The scheme of a typical PNR experiment is shown on Fig.1. The neutron beam of wavelength $\lambda$ is polarized by the polarizer, P (a magnetic supermirror [17] or polarized $^3$He gas chamber [18,19]), the transmission of which is dependent on the neutron spin direction. The polarized beam is then collimated by slits and directed to the sample (SM in Fig. 1) at an angle of incidence $\theta_1 \sim 10$ mrad. The neutron beam is scattered at an angle $\theta_2$. The polarization of the scattered beam is analyzed by an analyzer of polarization AP, the operating principle of which is the same as of P. The intensity transmitted through AP is incident on the detector D and is recorded. The beam polarization direction to the guide magnetic field in front of and behind the sample is periodically reversed by the spin-flippers (electromagnetic coils) $SF_1$ and $SF_2$.

The measured quantities in PNR experiments are intensities of the specular reflection ($\theta_2 = \theta_1$) and of the diffuse scattering ($\theta_2 \neq \theta_1$) at different angles of incidence and states of the spin-flippers (on-on, off-off, on-off and off-on). The primary data treatment procedure consists of taking into account the efficiencies of the polarizer, $P_p$, analyzer, $P_a$, spin flippers $P_{sf1(sf2)}$ and the background intensity. As a result of this processing one obtains the reflectivities $R^{\mu\eta}(Q)$ as a function of momentum transfer $Q$, normalized to the intensity of the incident beam. Here the indices $\mu$ and $\eta$ take values "+" or "-" and correspond to the projection of the neutron spin on the guide field before and after the scattering process, respectively. Experimental reflectivities are compared with the theoretical values. The latter are calculated varying the optical potential of the neutrons interaction with the investigated system.

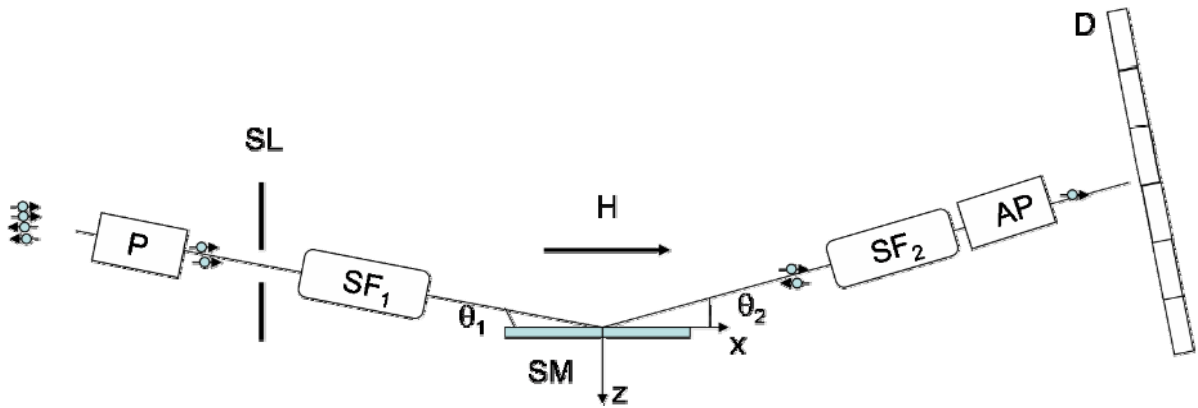

Fig 1. PNR experimental scheme. P – polarizer, SL – slit, $SF_{1(2)}$ – spin-flippers, H – external magnetic field, SM – sample, AP – analyzer of the polarization, D – detector

In the most general case the optical potential is a function of the three space coordinates and has nuclear and magnetic contributions:

$$V = 4\pi\rho_N + c\,\boldsymbol{\sigma}\mathbf{B}, \qquad (1)$$

where $\rho_N$ is the nuclear scattering length density (SLD), which depends on the nuclear density and coherent length scattering which is different for different atoms and isotopes [20]. The magnetic component is proportional to the scalar product of the Pauli matrix vector $\boldsymbol{\sigma}$ and the vector of magnetic induction in the medium $\mathbf{B}$. The scaling factor is $c = -\frac{2m}{\hbar^2}\mu_n = 2.91\cdot10^{-4}\,kOe\cdot nm^{-2}$, where $m$ and $\mu_n$ are the mass and the magnetic moment of the neutron, respectively, and $\hbar$ is the Planck constant. For a stratified medium, the optical potential depends on a single space coordinate, the one perpendicular to the layers. For this special case the solution of the Schrödinger equation can be obtained analytically [21,22,23]. Here we will present simple approximations to demonstrate the dependence of the reflectivity on the components of the optical potential in the different scattering channels. In particular, the amplitude of non-spin-flip specular scattering in the Born approximation can be written as:

$$r^{++}_{(--)}(Q) \approx \frac{1}{Q_z^2} \int \frac{\partial}{\partial z}\left[4\pi\rho_0(z) \pm cB_{\parallel}(z)\right]e^{iQ_z z}dz, \qquad (2)$$

where $Q_Z = 4\pi\sin\theta_1/\lambda$ is the plane-perpendicular momentum transfer component, $B_{\parallel}(z) = B(z)\cos\alpha(z)$ is the depth-profile of the parallel component of the magnetic induction on the neutron beam polarization vector ($\alpha$ is the angle between **B** and the polarization vector). From expression (2) it can be seen that amplitude $r^{++}$ ($r^{--}$) is a sum (difference) of Fourier transforms of the derivatives of the nuclear profile and $B_{\parallel}(z)$:

$$r^{++}_{(--)} = r_0 \pm r_{\parallel}^{mag}, \text{ where} \qquad (3)$$

$r_0, r_{\parallel}^{mag}$ are the reflection amplitudes due to a pure nuclear profile and $B_{\parallel}(z)$, respectively. For a weak magnetic system ($|cB|<<|4\pi\rho|$), consequently $r_0 >> r_{\parallel}^{mag}$, therefore the reflectivities $R^{++}_{(--)} \equiv \left|r^{++}_{(--)}\right|^2$ are primarily sensitive to the nuclear profile. In order to separate the magnetic contribution in the reflectivity it is convenient to use the so-called spin asymmetry $S(Q) = [R^{++} - R^{--}]/[R^{++} + R^{--}]$. From (2) and (3) it follows that the spin asymmetry is directly proportional to the Fourier transform of the gradient of $B_{\parallel}(z)$:

$$S(Q) \sim \int \frac{\partial B_{\parallel}(z)}{\partial z} e^{iQ_z z} dz \qquad (4)$$

The amplitudes of the spin-flip specular reflection can be calculated in the Distorted Wave Born approximation (DWBA) [24]:

$$r^{+-}_{(-+)}(Q) \approx \frac{c}{Q} \int \left(\psi^-(z)\right)^* B_\perp(z) \psi^+(z) dz, \text{ where} \qquad (5)$$

$\psi^{\pm}(z)$ are the neutron wave functions for the non-disturbed potential $4\pi\rho_N \pm B_{\parallel}(z)$, $B_\perp(z) = B(z)\sin\alpha(z)$ is the profile of the non-collinear component of the magnetic induction. At $Q_Z \to \infty$ the neutron wave-functions can be represented by plane waves and in that case expression (5) for the spin-flip amplitudes can also be represented by the Fourier transform of the non-collinear magnetic profile:

$$r^{+-}_{(-+)}(Q) \sim \frac{c}{Q} \int B_\perp(z) e^{iQ_z z} dz \qquad (6)$$

Thus, in the case of a one-dimensional optical potential, analysis of data allows us to recover the depth-profile of the nuclear potential and the vector of magnetic induction.

In the presence of in-plane inhomogeneities (interfaces roughness, magnetic domains, clusters, etc.) diffuse scattering will be observed. When the optical potential $\delta V(x,y)$ related to the inhomogeneities is small compared to the average potential, $<V(z)>$ the expression of the diffuse scattering amplitude can also be obtained in the DWBA:

$$r_{diff} \sim \int \left(\psi_2(z)\right)^* \delta V(x,y) \psi_1(z) dz, \text{ where} \qquad (7)$$

$\psi_{1(2)}(z)$ are the neutron wave functions inside the structure corresponding to the incoming and outgoing beams (at grazing angles of incidence $\theta_1$ and $\theta_2$, respectively). In (7) the spin components are omitted because their dependence on the components of the optical potential is similar to the case of specular reflection. Thus, the analysis of the intensity of diffuse scattering provides information on the distribution of nuclear and magnetic potential within the sample plane.

In most PNR experiment position sensitive detectors are used, which allow to record the specular and the diffuse scattering simultaneously. The typical duration of a wide $Q$-scan of a single SC/FM bilayer in four spin channels is 12-24 hours. In a typical one-week beam time it is possible to perform only 7-14 measurements at varying conditions (temperature, magnetic field, pressure). Moreover, irrespective of the length of the allocated beam time, the experimental background, and, consequently, the signal to background ratio can not be decreased below a certain limit. Therefore a certain level of the scattering intensity of a single interface should be maintained. For the above reasons we suggested to explore the waveguide enhancement of neutron standing waves [16]. The sample for such an experiment is prepared with the SC/FM bilayer placed between two layers with high nuclear SLD (Fig. 2). Upon certain conditions ($Q \equiv Q_{res}$), neutron waves propagating inside the SC/FM bilayer in both directions will interfere, which leads to the resonant increase of the amplitude of the wave function inside the investigated bilayer. As a consequence, intensities of the spin-flip and diffuse scattering (see expressions 5 and 7) will be significantly increased.

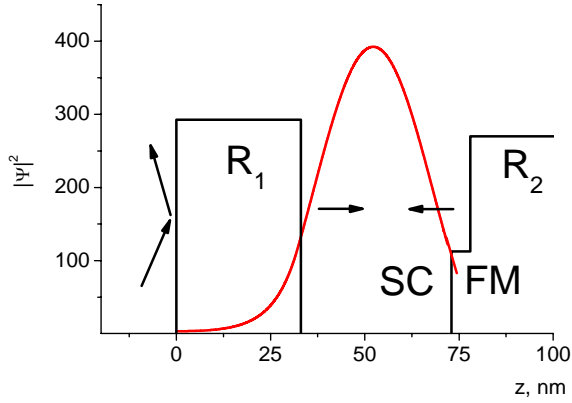

Fig. 2. Scheme of the waveguide enhancement. The nuclear SLD is shown by rectangular blocks and the SC/FM bilayer is inside the enhanced region [16]. Neutron waves propagating in opposite directions are shown by horizontal arrows. The profile of the density of neutrons $|\Psi|^2$ in the waveguide mode is shown in full (red) line.

In addition, the waveguide mode in the spin-flip specular reflection channel has several features [25] that allow us to gain additional information about the changes in the magnetic state of the SC/FM bilayer. These features are:

a) each waveguide mode in the spin-flip channel is represented by two adjacent resonant peaks (Fig. 3);

b) the "natural" width, $\Gamma$ of each resonance depends on the nuclear SLD and is extremely small ($\Gamma \sim 10^{-5} Q_{res}$);

c) the height, $H$ of each resonance depends on $B_\perp$, the component of the magnetic induction non-collinear to the neutron polarization;

d) the distance $2R_{res}$ between the peaks is linearly dependent on the magnitude of the magnetic induction in the SC/FM bilayer: $2R_{res} \sim \int |B(z)| dz$. Above $T_C$, when only the FM layer is magnetic, change of the peak separation may only be due to the change in the magnitude of the magnetization of the FM layer. Below $T_C$, the peak separation may change if additional magnetization appears in the SC layer. Figure 3 demonstrates the dependence of peak separation on the total magnetic moment of the SC/FM bilayer.

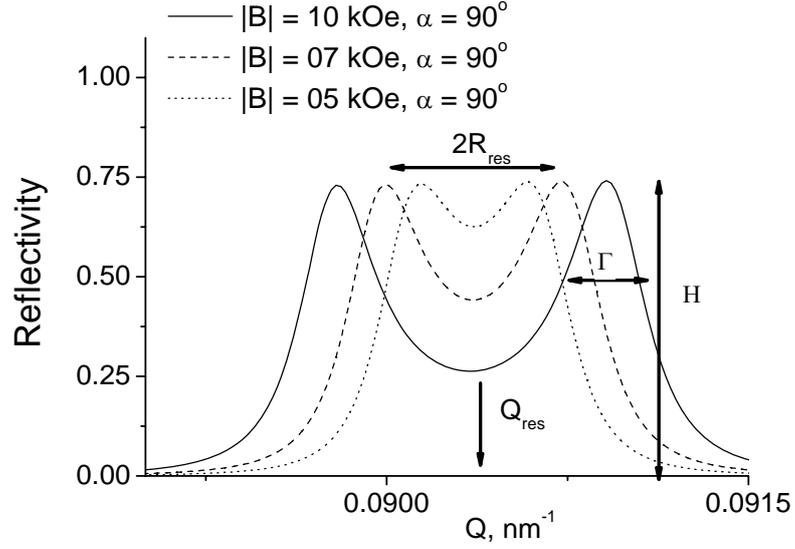

Fig. 3. The spin flip reflectivity in the vicinity of the waveguide mode for the different values of the magnitude of the magnetic induction in the SC/FM bilayer. In the figure: $Q_{res}$ and $R_{res}$, are the position of the resonance, and the peak separation, respectively. H and Γ are the height and the width of the resonances.

The experimentally measured intensity is the convolution of the intensity profile with the resolution function of the instrument. The latter can be represented by a Gaussian of width $\Delta Q$. $\Delta Q$ is the uncertainty in determining the momentum transfer that originates from the uncertainty in determining the wavelength, $\Delta\lambda$ and the angular divergence, $\Delta\theta$. The typical value of $\Delta Q \sim 10^{-2} Q_{res}$ is much larger than $2R_{res}$. As a consequence, the adjacent peaks in the spin-flip channel are not resolved. Representing the specular spin-flip intensity by a sum of two Gaussians of equal widths $\Gamma$, heights $H$ and a peak separation $2R_{res}$, the convolution with the resolution function limit gives a single Gaussian-like peak in the $\Delta Q \gg R_{res}$ limit. Its height $H^*$ and width $W^*$ can be calculated as:

$$H^* = \frac{2\Gamma}{\Delta Q} H \left(1 - \frac{R_{res}^2}{2\Delta Q^2}\right)$$

$$W^* = \Delta Q \left(1 + \frac{R_{res}^2}{2\Delta Q^2}\right) \qquad (8)$$

Since the peak parameter $H$ and $R_{res}$ are related to the components $B_\perp$ and $|B|$, the analysis of the height and width of the resonant peak (8) allows us to draw conclusions on the variation of the magnetic induction vector inside the SC/FM bilayer.

EXPERIMENT AND DISCUSSION

The sample with nominal composition of Cu(32nm)/V(40nm)/Fe(1nm)/MgO(001) and size of 20×10×2 mm$^3$ was prepared in the KFKI Research Institute for Particle and Nuclear Physics, Budapest by molecular beam epitaxy [26]. In order to create waveguide mode, as mentioned above the V(40nm)/Fe(1nm) SC/FM bilayer was prepared between the highly reflective MgO substrate and the Cu layer. The thickness of Cu was chosen to achieve the maximum enhancement of the specular spin-flip intensity [16]. Reflection high energy electron diffraction performed during growth confirmed an epitaxial growth of both the Fe and the V layers on MgO and Fe. Structural characterization of the sample [26] has shown high quality of the SC/FM interface. The rms roughness on MgO/Fe and Fe/V interfaces was estimated to be less than 0.6 nm. Magnetic properties of the FM layer were investigated by room temperature PNR and by SQUID magnetometry [26]. These measurements have confirmed a ferromagnetic state of the thin iron layer. The Curie temperature is estimated to be between 300K and 400K. The coercivity and saturation field at T=10K are $H_{coer}$ = 35 Oe and $H_{sat}$ =0.5 kOe, respectively. The saturation magnetization of the FM layer is $M_{sat}$ = 17.5 kGs which is 80% of the bulk value of the iron. The magnetization in remanence is only 10% below its saturation value. Such a slight decrease confirms that the magnetic state of the FM layer in remanence is close to its homogeneous state in saturation. The superconducting properties of the sample were checked by measuring the electrical resistivity of the sample with the four-terminal configuration [26]. According to these measurements, values for the in-plane upper critical field and the superconducting transition temperature are found to be $H_{C2}(0)$ = 15 kOe and $T_C(0)$ = 3.5K, respectively. The value of the superconducting coherence length was calculated from the $H_{C2}(T)$ dependence and gives $\xi$ = 9 ± 2 nm which is close to the values observed for similar Fe/V structures [27]. Thus, it is possible to draw the conclusion that the investigated FeV bilayer structure represents a smooth contact of a homogeneous 2D superconductor and a homogeneous ferromagnetic layer.

The low-temperature PNR measurements were conducted at the ADAM (Institute of Laue-Langevin, France) [28] and the V6 (Helmholtz-Zentrum für Materialien und Energie, Germany) [29] reflectometers. The principal parameters of the instruments and the conditions of the experiments are presented in Table 1.

Table 1. Parameters of the experimental setups. Here $\lambda$ is the neutron wavelength, $P_p$ is the beam polarization and $dQ/Q_{res}$ is the experimental resolution at the waveguide resonance position.

| Instrument | $\lambda$ [Å] | $P_p$, % | $dQ/Q_{res}$,% | Analyzer | Cryostat |
|---|---|---|---|---|---|
| ADAM | 4.4 ± 0.1 | 95 | 5.3 | $^3$He cell | continuous flow cryostat, stability 0.05K |
| V6 | 4.7 ± 0.1 | 93 | 5.8 | Si/Fe magnetic supermirror | closed cycle cryostat, stability 0.1K |

In order to use the waveguide regime described above it is necessary to increase $B_\perp$, the non-collinear component of the magnetic induction. To do so we magnetized sample along the short edge of the substrate at $T$ = 30K and then cooled the sample down to $T$ = 1.6K in zero field. Then a small guide field was applied along the long side of the sample and the reflectivity scans were started. The applied guide field of $H \approx$ 20 Oe was lower than $H_{coer}$, but sufficient to keep the neutron polarization parallel to **H**. Thus the described experiments can be considered to have been performed in practically zero field. The appearance of induced magnetization in the SC layer, therefore, can not be due to a direct effect of the field on the SC (like Meissner effect, or vortex state or such).

In Figure 4 the specular reflectivities $R^{++}(Q)$ and $R^{+-}(Q)$ measured on ADAM at $T$ = 5K are presented. The non-spin-flip reflectivity $R^{++}(Q)$ is characterized by the region of total external reflection $Q < Q_{cr}$ = 0.17 nm$^{-1}$, the Kiessig oscillations above $Q_{cr}$ and a waveguide resonance dip at $Q_{res}$ =0.1nm$^{-1}$. The spin-flip reflectivity $R^{+-}(Q)$ is characterized by the presence of the waveguide resonance peak at $Q_{res}$ with the intensity of about 10% of the intensity of the direct beam.

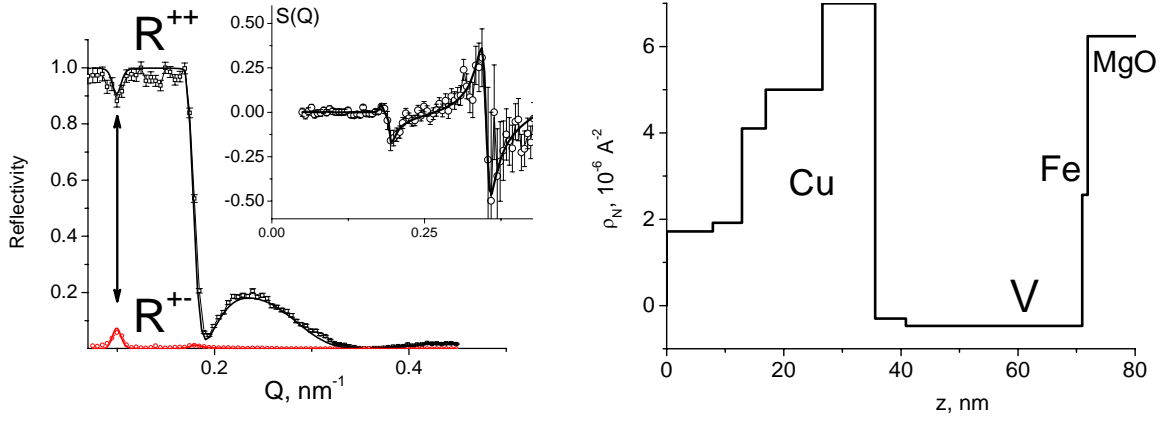

Fig 4. a) Experimental (dots) and model (lines) reflectivity curves in different channels measured at ADAM at $T = 5K$ Arrow shows the position of the waveguide resonance. Inset: spin asymmetry. b) Model potential of interaction obtained as a result of the fit shown in a).

We fitted a theoretical model potential set to the PNR data in the same way as described in [16]. The results of the fit, the nuclear SLD profile is shown in Fig. 4-b. The profile qualitatively agrees with the results presented in [26] obtained by different methods. The thickness of the vanadium and the copper layers are close to nominal. Nuclear SLD of Cu, V and MgO are also close to the literature values. As expected, the fit was insensitive to the parameters of the iron layer in this $Q$-range. A simultaneous fit of $R^{+-}(Q)$ and $S(Q)$ provided the magnitude $|M|$ and angle $\alpha$ of the FM magnetization vector above $T_C$. The magnitude is $|M| = 14 \pm 1$ kOe which is in good agreement with SQUID data of $15.5 \pm 0.5$ kOe. According to the best fit, the vector of the FM magnetization is tilted relative to the external field at an angle of $\alpha = 45 \pm 5°$. It is very likely that external field comparable to the $H_{coer}$ caused the in-plane rotation of the magnetization from the expected 90° to the observed 45°.

The waveguide enhancement allowed us to measure a detailed temperature dependence of the spin-flip and diffuse scattering intensities in the $Q$-range of the waveguide mode near $T_C$. The temperature scans were performed by rising the temperature from 1.6 K to 6 K with a step of 0.2 K for $T \leq T_C$ and a step size of 0.5 K for $T > T_C$. Data acquisition time and relative statistical error at each temperature was 30 to 60 minutes and 1-3%, respectively. As a result, detailed information about variation of the magnetic state of SC/FM bilayer was collected in a wide temperature range within a relatively short time (about 12 hours). In Fig. 5, the temperature dependence of height (a) and width (b) of the $R^{+-}$ peak measured at the V6 reflectometer is presented. One can separate two regions below $T_C$. In the $T = [1.5 \div 2.6]$ K region the peak maximum (width) is almost constant and lower (higher) by 20% (14%) as compared to the corresponding parameter above $T_C$. Upon further increasing the temperature, a monotonous increase (decrease) of the height (width) of the composite peak is observed to values corresponding to the normal state of the sample at $T > T_C$.

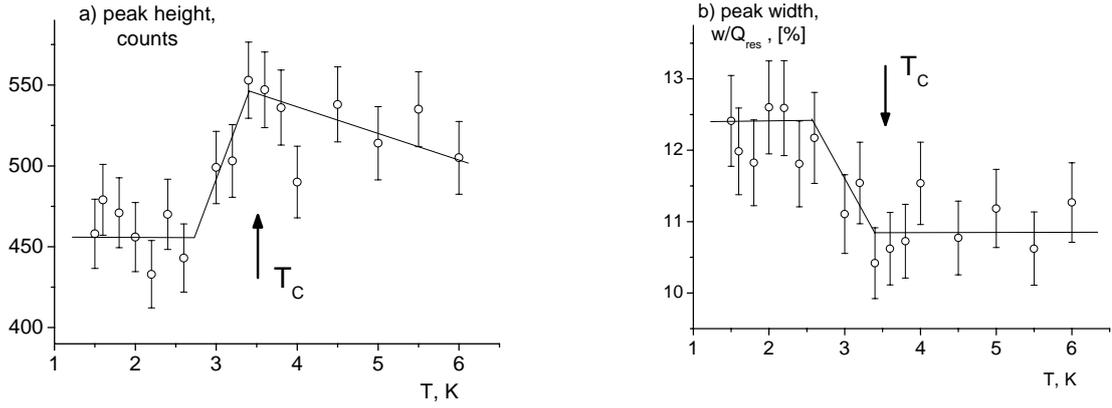

Fig. 5. The maximum (a) and the width (b) of the measured $R^{+-}$ resonant spin-flip peak as a function of the temperature. The solid lines are only guides for the eye.

In addition to the specular scattering, the intensity of the diffuse scattering at the waveguide mode was measured. The temperature dependence of the diffuse scattering intensity in the (+ +) channel integrated over scattering angles $\theta_2 = [6 \div 60]$ mrad measured on the V6 instrument is shown in Fig. 6. In the $T = (2.5 \div 3.5)$ K temperature range the diffuse scattering intensity increases (by 5% with a statistical error of 1.5 %) as compared to the values in the $T < 2.5$ K temperature range. The temperature evolution of the diffuse scattering in the (+ -) channel (not shown) correlates with the dependence of the specular (+ -) spin-flip scattering intensity. This allows us to prefer an interpretation of the changes of the diffuse scattering in this channel by changes in the depth-profile of the magnetic induction rather than otherwise possible changes in the domain state.

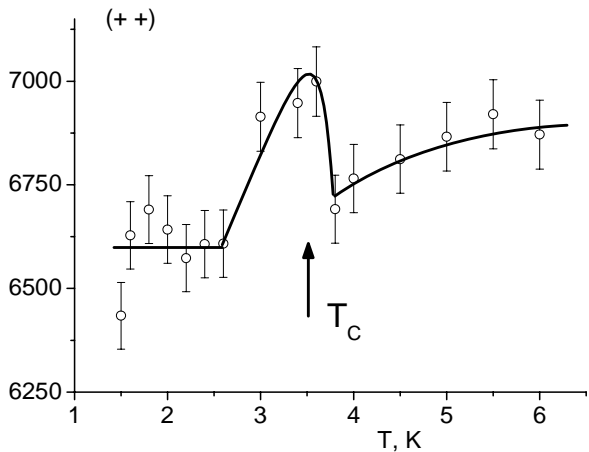

Fig. 6. Temperature dependence of integrated intensity of the diffuse scattering in the (+ +) channel. The solid line is a guide for the eye.

Preliminary analysis of the waveguide enhanced PNR data of Cu(32nm)/V(40nm)/Fe(1nm)/MgO(001) near the waveguide resonance is indicative of the appearance of a specific magnetic state of the SC/FM bilayer in the $T = [1.5 \div 2.6]$ K temperature range: This state is different from that of the bilayer above $T_C$. Upon further increasing the temperature, a gradual decay of this state is observed which is accompanied by an increase of diffuse scattering. The changes in the magnetic state of the bilayer are only observed below $T_C$, therefore we conclude that it originates from the proximity of the SC and the FM layers.

Quantitatively, the observation below $T < 2.5$K can be described by two different models. One may assume that the change of the magnetization takes place in the FM layer. Then below 2.5K the magnitude of the magnetization must increase to $|M| = 19 \pm 1$ kOe which is higher than the saturation value found by SQUID. Moreover, in order to quantitatively describe the decrease of the peak height we have to introduce the rotation of the FM magnetization towards the external field by 5-10 degrees from the initial value. Presently we can't suggest a physical phenomenon to be responsible for a simultaneous growth and rotation of the magnetization vector. The alternative model to quantitatively explain the change of the magnetization below $T = 2.5$ K is the appearance of an additional magnetization in the SC layer. In this model the peak broadening can be assigned to the additional *positive* magnetic moment in SC layer with a value of +5 kGs×nm. This means that assuming the thickness of a new magnetic sub-layer inside the SC layer to be 10 nm (close to $\xi$), the magnetization of this magnetic sub-layer is +0.5 kGs. In order to quantitatively describe the decrease of the height of the waveguide peak we should require tilting angle $\alpha$ for the magnetization of this sub-layer to be about $\alpha \sim 20^o$. The appearance of this magnetic moment in the SC lazer may well be due to the proximity of the FM layer. This is indicated by the magnitude of the effect and the character of the temperature behavior similar to the theoretical predictions.

CONCLUSION

A single SC/FM bilayer with composition Cu(32nm)/V(40 nm)/Fe(1 nm)/MgO has been investigated by waveguide enhanced polarized neutron reflectometry. The main results of the present work can be summarized as follows. In the $T = [0.4 \div 0.7]$ $T_C$ temperature range a new magnetic state was observed characterized by a 14% broadening and 20% height reduction of the waveguide mode spin-flip peak. Upon increasing the temperature from 0.7 $T_C$ to $T_C$, a gradual decay of this state was observed, accompanied by the 5% growth of the diffuse scattering intensity. This behavior can be explained in a natural way by the polarization of the superconducting electrons upon the SC transition, i.e. an appearance of additional induced magnetization within the SC, due to the proximity of the FM layer. This is supported by the temperature evolution and length scale of the penetration of the new state. However, some complicated scenarios without SC electron polarization, e.g. a simultaneous increase and rotation of the FM layer magnetization can not be entirely excluded at present.

ACKNOWLEDGEMENTS

The authors would like to thank A. Paul and D. Wallacher (Helmholtz-Zentrum für Materialien und Energie) for assistance in measurements PNR at V6 instrument. This work was partially supported by the Russian Foundation for Basic Research (grant № 09-02-00566), the bilateral JINR-HAS project (EAI-2009/002) and the National Office for Research and Technology of Hungary (NAP-VENEUS project). Yu. Khaydukov would like to acknowledge the support of the Foundation for Assistance to Small Innovative Enterprises of the Russian Federation (grant № 8455 UMNIK-08-3). B. Nagy and Yu. Khaydukov would like to thank the generous mobility grant of HZB for the experimental session.